\documentclass[runningheads]{llncs}
\usepackage[T1]{fontenc}
\usepackage{amsmath}
\usepackage{cite}
\usepackage{todonotes}

\usepackage{graphicx}

\usepackage{xcolor}         
\usepackage{hyperref}       

\hypersetup{
    colorlinks=true,        
    linkcolor=magenta,      
    citecolor=magenta,      
    filecolor=magenta,      
    urlcolor=magenta        
}
\begin{document}
\title{Mammographic Breast Positioning Assessment via Deep Learning}


%
%


\author{
Toygar Tanyel\inst{1} \and
Nurper Denizoglu\inst{2} \and
Mustafa Ege Seker\inst{3} \and
Deniz Alis\inst{4} \and \\
Esma Cerekci\inst{5} \and
Ercan Karaarslan\inst{4} \and
Erkin Aribal\inst{4} \and 
Ilkay Oksuz\inst{6}
}

\institute{
Istanbul Technical University, Biomedical Engineering Graduate Program \and 
Acibadem Healthcare Group, Department of Radiology \and
Acibadem Mehmet Ali Aydinlar University, School of Medicine \and
Acibadem Mehmet Ali Aydinlar University, School of Medicine, Dept. of Radiology \and
Sisli Hamidiye Etfal Training and Research Hospital \and
Istanbul Technical University, Department of Computer Engineering \\
\email{\{tanyel23, oksuzilkay\}@itu.edu.tr}
}


%
\authorrunning{Tanyel et al.}
%
%
\maketitle             

\begin{abstract}
Breast cancer remains a leading cause of cancer-related deaths among women worldwide, with mammography screening as the most effective method for the early detection. Ensuring proper positioning in mammography is critical, as poor positioning can lead to diagnostic errors, increased patient stress, and higher costs due to recalls. Despite advancements in deep learning (DL) for breast cancer diagnostics, limited focus has been given to evaluating mammography positioning. This paper introduces a novel DL methodology to quantitatively assess mammogram positioning quality, specifically in mediolateral oblique (MLO) views using attention and coordinate convolution modules. Our method identifies key anatomical landmarks, such as the nipple and pectoralis muscle, and automatically draws a posterior nipple line (PNL), offering robust and inherently explainable alternative to well-known classification and regression-based approaches. We compare the performance of proposed methodology with various regression and classification-based models. The CoordAtt UNet model achieved the highest accuracy of 88.63\% $\pm$ 2.84 and specificity of 90.25\% $\pm$ 4.04, along with a noteworthy sensitivity of 86.04\% $\pm$ 3.41. In landmark detection, the same model also recorded the lowest mean errors in key anatomical points and the smallest angular error of 2.42 degrees. Our results indicate that models incorporating attention mechanisms and CoordConv module increase the accuracy in classifying breast positioning quality and detecting anatomical landmarks. Furthermore, we make the labels and source codes available to the community to initiate an open research area for mammography, accessible at \href{https://github.com/tanyelai/deep-breast-positioning}{https://github.com/tanyelai/deep-breast-positioning}.

\end{abstract}
\section{Introduction}
Breast cancer remains the most common cancer and leading cause of cancer-related deaths among women globally \cite{globalcancer}. Mammography screening is the most effective method for early detection, significantly reducing mortality rates \cite{magnus2011effectiveness}. Thus, many countries have adopted national screening programs \cite{duffy2002impact}.

A standard mammogram includes craniocaudal (CC) and mediolateral oblique (MLO) views, with the MLO view being crucial as it captures nearly the entire breast tissue, especially the upper quadrant where cancer frequently occurs. Proper positioning in mammography is vital, as poor positioning can lead to diagnostic errors and necessitate repeat exams, increasing costs and causing stress for patients \cite{fda2016, mackenzie2016, feig2002, gurdemir2012}. There is a pressing need for automated systems that can instantly evaluate the quality of mammogram positioning, allowing technologists to take immediate corrective action if necessary.

Recent advancements in deep learning (DL) have shown promising results in breast cancer diagnostics, often matching or surpassing radiologists in accuracy \cite{geras2019artificial, rodriguez2019stand}. However, less focus has been given to using DL for assessing mammography positioning. This gap presents an opportunity to improve the evaluation process right after image acquisition. Traditionally, studies have relied on classification-based DL approaches involving qualitative expert assessments \cite{brahim2022automated, watanabe2023quality} or by dividing related tasks into separate regression processes \cite{gupta2020deep}, which can introduce additional complexity and affect the explainability and objectivity of the results.

Our work introduces a novel deep learning methodology that quantitatively evaluates image positioning quality in MLO views. By identifying key anatomical features such as the nipple and pectoralis muscle, and automatically drawing a perpendicular posterior nipple line (PNL) to the pectoralis muscle or film edge, our methodology provides a robust and superior alternative to traditional classification-based approaches. We demonstrate the effectiveness of our method on existing models, showcasing its potential to enhance mammography positioning assessments.


\section{Materials} 
In this section we provide information on the dataset and the ground truth criteria for correct MLO positioning.

\subsection{Study Sample}

We used the VinDr Mammography dataset \cite{nguyen2023vindr}, an open-access collection of 5000 exams from two hospitals in Vietnam (2018-2020). From this, we selected 1000 exams, each with two MLO view mammograms from both breasts, totaling 2000 images. Exams were split into training, validation, and testing sets with an 80\%/10\%/10\% split, ensuring a balanced representation of clinical outcomes. According to the PNL criteria, MLO-view positioning in the datasets was classified as 967 good and 633 poor for training, 108 good and 92 poor for validation, and 123 good and 77 poor for testing.

\subsection{Image Positioning Quality Criterion}

Several international systems assess the quality of mammography images in MLO views based on criteria like the angle, width, and length of the pectoral muscle, its border angulation, and the distance between the pectoral muscle and the nipple. The primary goal is to ensure maximum breast tissue coverage. Some criteria, such as the distance from the pectoral muscle to the nipple, are subjective and impractical \cite{spuur2011mammography}. The angle and dimensions of the pectoral muscle lack universal standards. A consistent criterion is that the PNL, drawn from the nipple to the pectoralis muscle at a right angle, intersects the pectoralis muscle. This method is endorsed by the American College of Radiology and the Royal Australian and New Zealand College of Radiologists \cite{hendrick1999, australian2001, rancr2002, wilson2011, spuur2011mammography} and is adopted as our study's reference standard.

\subsection{Ground Truthing Process}
Ground truth annotations were performed by a board-certified breast radiologist (N.D.) with over five years of experience in breast imaging. The radiologist used a specialized workstation, featuring a browser-based annotation tool (https://matrix.md.ai) and a 6-megapixel diagnostic monitor (Radiforce RX 660, EIZO), to annotate mammograms. All mammograms were examined in the Digital Imaging and Communications in Medicine (DICOM) format. The radiologist marked the nipple and pectoralis muscle line on MLO views.


\section{Methods} 

In this section we provide details of our pre-processing operation, loss function, model architecture and experimental setup.

\subsection{Pre-processing Steps}
The pre-processing steps for mammography images are designed to prepare the data for analysis while preserving anatomical features and spatial relationships. Initially, the midpoint of the nipple bounding boxes and the endpoints of the pectoralis muscle are extracted, yielding three critical landmarks for orientation and scale adjustments. Next, the endpoints of the pectoralis muscle are standardized by extending each line to the image boundary with a 10-pixel margin to minimize variability from radiologists' arbitrary line terminations. Significant breast regions are then isolated by removing extensive black pixel areas around the periphery and below the breast. This involves thresholding the image to create a binary version, applying morphological opening, labeling connected regions, and cropping the image to the bounding box of the largest region. Zero-padding is applied to make the images square, preventing distortion during resizing and maintaining uniformity across the dataset. Subsequently, all images are resized to $512 \times 512$ pixels to facilitate computational efficiency and model training while preserving necessary detail. 

\subsection{Landmark-Aware Wing Loss} 
Landmark-Aware Wing Loss is tailored to improve the model's accuracy in predicting landmark coordinates. It employs a piecewise function that combines the Wing Loss's sensitivity to small errors with a linear part to moderate the response to larger errors. The Wing Loss \cite{feng2018wing} formula is given by:
\begin{equation}
\mathcal{L}_{Wing}(y) = 
\begin{cases} 
w \cdot \log(1 + \frac{|y|}{\epsilon}), & \text{if } |y| < w \\
|y| - C, & \text{otherwise}
\end{cases}
\end{equation}
where $y$ represents the absolute error between the predicted and target coordinates, $w$ is the parameter that defines the width of the non-linear region, $\epsilon$ controls the curvature within this region, and $C$ is a continuity constant, defined as $C = w - w \cdot \log(1 + \frac{w}{\epsilon})$.

Practically, the $\mathcal{L}_{LAW}(y)$ is calculated for each coordinate of the landmarks, leading to a comprehensive loss for each landmark by summing up the mean losses of the coordinates, i.e., mean of x and y, expressed as:
\begin{equation}
\mathcal{L}_{LAW} =  \alpha \cdot \mathcal{L}_{Wing(L1)} + \beta \cdot \mathcal{L}_{Wing(L2)} + \gamma \cdot \mathcal{L}_{Wing(L3)}
\end{equation}
where $\alpha$, $\beta$, and $\gamma$ are the weights for the landmarks.

\subsection{Model Architectures and Techniques for Landmark Detection}

\begin{figure}[htb!]
\begin{center}
\resizebox{\textwidth}{!}{
\includegraphics[]{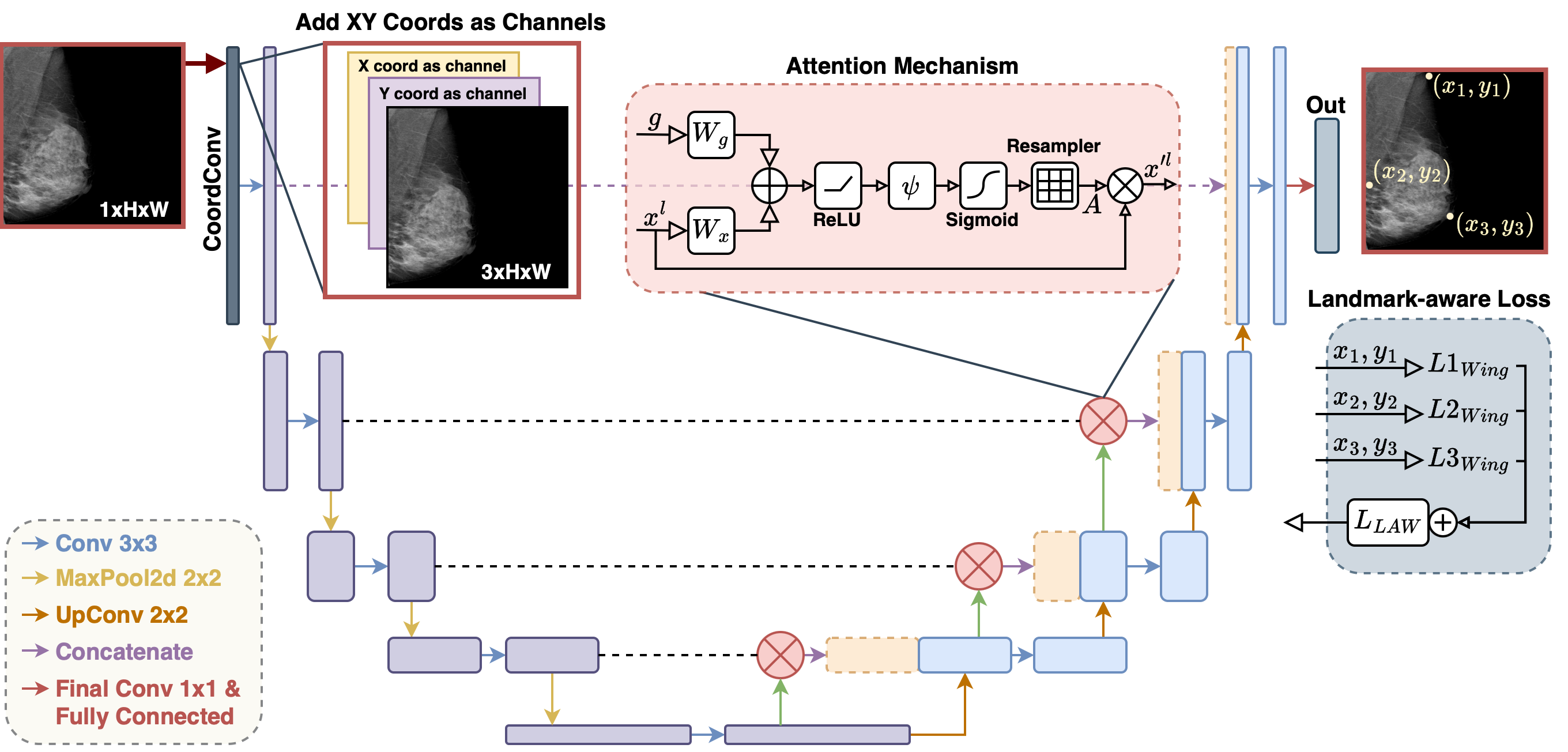}}
\caption{Illustration of concepts utilized in this study as part of an ablation study. At the input layer, a single-channel grayscale mammogram is augmented to a three-channel image by introducing two additional channels that encode the X and Y spatial coordinates of each pixel. The attention mechanism refines features, and skip connections preserve spatial information. The final layer outputs landmark coordinates, optimized using landmark-aware wing loss.} \label{Fig2}
\end{center}
\end{figure}

We used U-Net as the backbone with coordinate convolution module (CoordConv) and attention mechanisms for landmark detection (Fig. \ref{Fig2}). CoordConv and attention refine feature maps, improving spatial information. We also used ResNeXt50 as the backbone for landmark regression, adjusting it for single-channel input and comparing its classification and regression results. This section details these components and the landmark regression process.

\subsubsection{CoordConv Integration}

To improve spatial awareness, we replaced the initial convolutional layer with a CoordConv layer, integrating spatial coordinates into the convolution operation. CoordConv augments the input by adding normalized spatial coordinates across height (H) and width (W), improving the model's ability to learn spatial hierarchies:

\[
\mathcal{CC}(I) = \mathcal{R} \left( \mathcal{B} \left( \mathcal{C} \left( \mathcal{A}(I, \{x, y\}) \right) \right) \right),
\]

where $\mathcal{A}(I, \{x, y\})$ adds coordinate channels to input tensor $I$. The x-coordinates $\{x\}$ and y-coordinates $\{y\}$ are normalized in the range [0, 1], calculated as:
    \[
    x_j = \frac{j}{W-1}, \quad y_i = \frac{i}{H-1},
    \]
    for each pixel $(i, j)$ in the feature map. These coordinates are repeated across the batch size and stacked to form two new channels appended to $I$.

$\mathcal{C}$ performs convolution with these channels, and $\mathcal{B}$ and $\mathcal{R}$ represent batch normalization and ReLU activation.

\subsubsection{Attention Mechanism}

Attention mechanisms in the U-Net model selectively focus on important features. The attention block integrates features from both encoder and decoder paths by computing attention coefficients (\(\psi\)) using the following formulations: \(g = W_{\text{gate}} \ast G\) and \(x = W_{x} \ast X\). Here, \(W_{\text{gate}}\) and \(W_{x}\) are convolutional filters, with \(G\) representing the gating signal from the decoder and \(X\) the feature map from the encoder. The attention coefficients \(\psi\) are computed as \(\psi = \sigma(\text{ReLU}(g + x))\), where \(\sigma\) denotes the sigmoid activation. The attended output \(A\) is then calculated as \(A = X \cdot \psi\). This mechanism ensures that only the most relevant features are propagated through the network to improve the precision of the output.



\subsubsection{ResNeXt50}

We used ResNeXt50, a 50-layer deep residual network with a cardinality of 32, featuring grouped convolutions for enhanced feature learning. For landmark regression, we modified ResNeXt50 to accept single-channel input and output the required landmark coordinates. Additionally, we used the raw classification model to compare image-level classification with regression results, evaluating differences in accuracy and robustness, and demonstrating the model's versatility and effectiveness.

\subsection{Evaluation}
\label{sec:evaluation}

The evaluation checks our model's accuracy in predicting the nipple and endpoints of the pectoral muscle line. We verify if the perpendicular intersection from the nipple to the pectoral muscle (i.e., PNL) falls within the image boundaries to indicate image quality. We also measure the angular error between the original and predicted pectoral muscle lines. Additionally, we compute accuracy, sensitivity, and specificity for the model's decisions. Euclidean distance between predicted and original landmarks in millimeters is calculated, ensuring real-world measurement accuracy. The angle error is normalized to 0-180 degrees to represent the deviation from vertical. Performance metrics are based on image quality classification, with accuracy as the proportion of correct predictions, sensitivity as the proportion of correctly identified bad quality images, and specificity as the proportion of correctly identified good quality images.

\subsection{Experimental Setup}

\subsubsection{Regression}
We employed the R-ResNeXt50, UNet, Attention UNet, and CoordAtt UNet models, each configured with a single input channel and six output channels corresponding to the $x,y$ coordinates of three landmarks. Training was conducted on an NVIDIA L4 GPU for 300 epochs. The Adam optimizer was used with an initial learning rate of $1 \times 10^{-4}$, dynamically adjusted by a CyclicLR scheduler oscillating between $1 \times 10^{-5}$ and $5 \times 10^{-4}$ using a triangular policy without cycle momentum. The loss function integrated Wing Loss ($w=3$, $\epsilon=1.5$) and additional parameters ($\alpha=1.0$, $\beta=1.0$, $\gamma=1.0$) to capture both precision and geometric intricacies in landmark detection. The model with the lowest validation loss was preserved for subsequent evaluation. As an exception, the R-ResNeXt50 model was initially fine-tuned with pretrained ImageNet weights using a batch size of 8 for 150 epochs.

\subsubsection{Classification}
For classification, we utilized a ResNeXt50 model to classify images into two positioning quality classes. Training was conducted on an NVIDIA L4 GPU with a batch size of 8 for 30 epochs, fine-tuning the model with pretrained ImageNet weights. The same optimizer and learning rate progression as in the regression setup were applied. Categorical Cross-Entropy Loss was used for loss calculations. The best-performing model, determined by a range of metrics, was preserved for subsequent evaluation.

\section{Results}

In this section, we provide results on landmark detection and binary image quality assessment.

\textbf{Models’ Performance on Landmark Detection.} 

Table \ref{tab2} details models' performance on landmark detection, focusing on distance errors. Direct landmark regression with ResNeXt50 (R-ResNeXt50)  and vanilla UNet regression (UNet) showed higher errors for pectoral and nipple landmark detection. Attention UNet considerably improved performance both in terms of median errors, and an angular errors. CoordAtt UNet outperformed others with mean errors of 4.99mm (Perp), 5.62mm (Pec1), 6.49mm (Pec2), 2.97mm (nipple), and the smallest angular error of 2.42 degrees.

\renewcommand{\arraystretch}{1.3}
\begin{table}[htb!]
\centering\caption{Distance errors in millimeters (mm), presented as mean ($\mu$), standard deviation ($\sigma$) and median ($x\sim$) to mitigate the influence of challenging cases (primarily due to subjectivity of the task). Perp: Perpendicular intersection error for the line drawn from the nipple to the pectoral line. Pec1: Lower endpoint of the pectoral muscle line error. Pec2: Upper endpoint of the pectoral muscle line error. Nipple: Nipple location error. Angular: Angular difference between the predicted and original pectoral muscle line.}
\resizebox{\textwidth}{!}{\label{tab2}%
\begin{tabular}{cccccccccccccccc}
\cline{2-16}
\multicolumn{1}{l}{}        & \multicolumn{3}{c}{\textbf{Perp}}                                   & \multicolumn{3}{c}{\textbf{Pec1}}                                   & \multicolumn{3}{c}{\textbf{Pec2}}                                   & \multicolumn{3}{c}{\textbf{Nipple}}                                 & \multicolumn{3}{c}{\textbf{Angular}}           \\ \hline
\multicolumn{1}{c|}{\textbf{Models}} & \textbf{$\mu$}    & \textbf{$\sigma$} & \multicolumn{1}{c|}{\textbf{$x$$\sim$}} & \textbf{$\mu$}    & \textbf{$\sigma$} & \multicolumn{1}{c|}{\textbf{$x$$\sim$}} & \textbf{$\mu$}    & \textbf{$\sigma$} & \multicolumn{1}{c|}{\textbf{$x$$\sim$}} & \textbf{$\mu$}    & \textbf{$\sigma$} & \multicolumn{1}{c|}{\textbf{$x$$\sim$}} & \textbf{$\mu$}    & \textbf{$\sigma$} & \textbf{$x$$\sim$} \\ \hline
R-ResNeXt50                 & 7.13          & 4.23       & 6.49                                   & 7.33          & 6.01       & 5.24                                   & 7.93          & 7          & 6.2                                    & 4.63          & 1.99       & 4.45                                   & 2.71          & 2.44       & 1.96              \\
UNet                        & 9.62          & 7.86       & 8.03                                   & 8.19          & 6.89       & 6.01                                   & 14.01         & 14.01      & 10.9                                   & 6.8           & 5.25       & 5.72                                   & 3.52          & 3.15       & 2.66              \\ \hline
Attention UNet              & 5.12          & 5.04       & \textbf{3.56}                          & 6.01          & 5.87       & \textbf{4.03}                          & 6.94          & 8.25       & \textbf{3.95}                          & 2.98          & 2.4        & 2.52                                   & 2.58          & 2.73       & 1.81              \\
CoordAtt UNet               & \textbf{4.99} & 4.88       & 3.82                                   & \textbf{5.62} & 5.29       & 4.14                                   & \textbf{6.49} & 7.37       & 4.26                                   & \textbf{2.97} & 2.46       & \textbf{2.45}                          & \textbf{2.42} & 2.56       & \textbf{1.75}     \\ \hline
\end{tabular}}
\end{table}

\begin{figure}[htb!]
\begin{center}
\resizebox{\textwidth}{!}{
\includegraphics[]{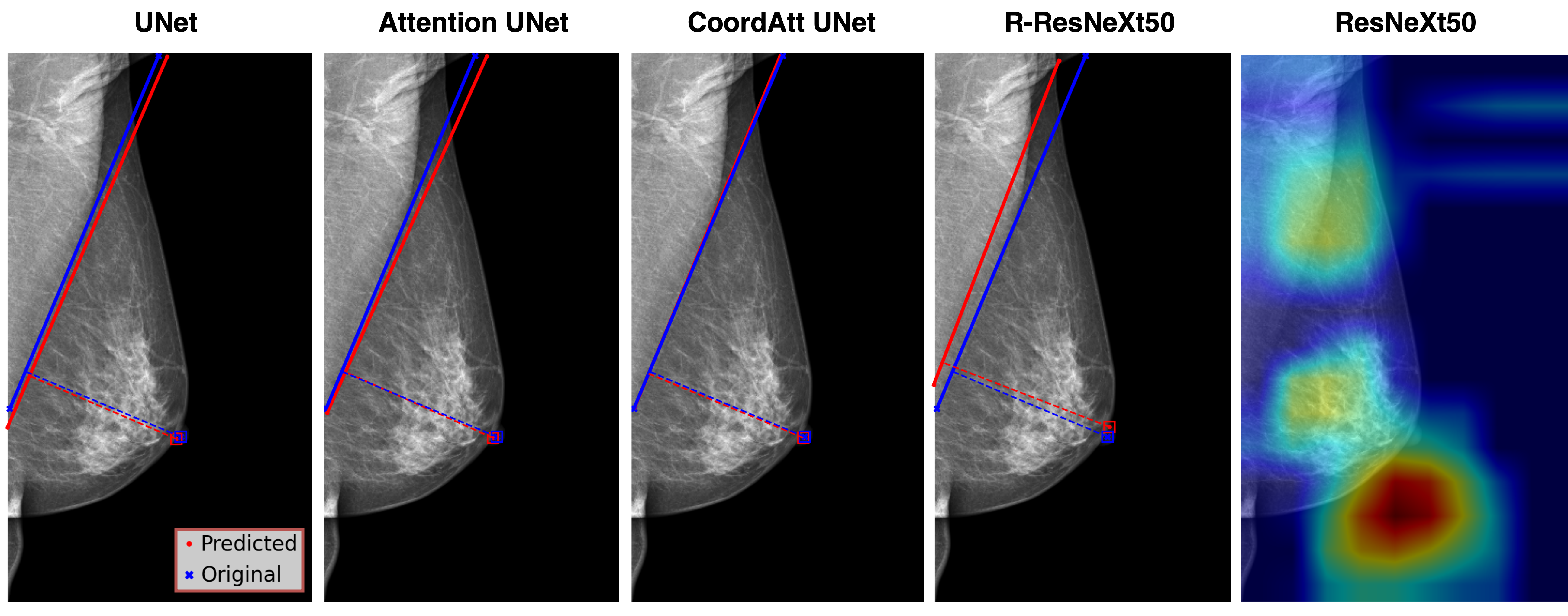}}
\caption{Comparison of predicted (red) versus original (blue) landmarks in mammograms using different models: UNet, Attention UNet, CoordAtt UNet, R-ResNeXt50, and ResNeXt50. The rightmost column shows heatmaps for ResNeXt50.} \label{Fig3}
\end{center}
\end{figure}

\textbf{Models’ Performance on Breast Positioning Labels.} 

The models' performance on breast positioning labels is summarized in Table \ref{tab1}. The raw ResNeXt50 model used for binary classification without landmark regression achieved the lowest performance measures. Addition of landmark regression and rule-based binary classification (R-ResNeXt50), improved the performance dramatically. Vanilla UNet regression of the landmark points showed poor performance, where Attention block addition (Attention UNet) outperformed previous models. Addition of coordinate points (CoordAtt UNet) achieved comparable performance to Attention UNet showcasing superior performance in accuracy and specificity.

\renewcommand{\arraystretch}{1.3}
\begin{table}[htb!]
\centering\caption{Test results on automatically generated quality labels extracted from radiologists' label drawings. The raw ResNeXt50 model was trained for binary classification based on image-level labels. The R-ResNeXt50 model had its last layer modified to function as a landmark regressor, predicting coordinates and overall positioning quality, similar to our proposed pipeline. Results are presented as the mean $\pm$ standard deviation of 5 different training runs.}
\resizebox{0.7\textwidth}{!}{\label{tab1}%
\begin{tabular}{cccc}
\hline
\textbf{Model} & \textbf{Accuracy}     & \textbf{Specificity}  & \textbf{Sensitivity}  \\ \hline
ResNeXt50      & 73.7 $\pm$ 3.35           & 76.91 $\pm$ 6.26          & 68.57 $\pm$ 11.41         \\ 
R-ResNeXt50     & 82.3 $\pm$ 5.03           & 81.42 $\pm$ 12.34         & 83.38 $\pm$ 10.49         \\
UNet           & 70.63 $\pm$ 1.49          & 78.46 $\pm$ 1.56          & 58.12 $\pm$ 2.68          \\ \hline
Attention UNet & 88.2 $\pm$ 2.51           & 88.62 $\pm$ 4.11          & \textbf{87.53 $\pm$ 3.51} \\
CoordAtt UNet  & \textbf{88.63 $\pm$ 2.84} & \textbf{90.25 $\pm$ 4.04} & 86.04 $\pm$ 3.41  \\  \hline      
\end{tabular}}
\end{table}

\section{Discussion}

In this study, we presented a novel deep learning methodology for assessing the quality of mammogram positioning, focusing on the MLO views. Our method quantitatively evaluates image positioning by identifying key anatomical features and drawing a perpendicular PNL, providing a robust alternative to traditional classification-based approaches. The evaluation of various deep learning models, including ResNeXt50, UNet, Attention UNet, and CoordAtt UNet, demonstrated considerable improvements in accuracy, specificity, and sensitivity. Notably, the CoordAtt UNet model achieved the highest performance, highlighting the effectiveness of incorporating attention mechanisms and CoordConv module (Fig. \ref{Fig3}). This study addresses a critical unmet need in mammography screening, offering an automated, objective, and explainable solution for assessing breast positioning quality, which is crucial for accurate breast cancer diagnosis.

Despite the promising results, several limitations must be acknowledged. Our study focused exclusively on MLO views, which, while comprehensive, do not cover all diagnostic perspectives. Future work will extend the model to include CC views to provide a more holistic evaluation of mammogram positioning. Additionally, our primary criterion for evaluating positioning quality was the PNL. While robust for MLO views, the models' effectiveness might be limited when considering other criteria, such as the angle and shape of the pectoral muscle. Future studies will aim to incorporate these additional criteria to improve the models' versatility. The clinical impact of this research is significant, as it paves the way for more reliable and efficient mammography screening, ultimately improving early breast cancer detection and patient outcomes.

\bibliographystyle{splncs04}
\bibliography{mybib}

\begin{thebibliography}{10}
\providecommand{\url}[1]{\texttt{#1}}
\providecommand{\urlprefix}{URL }
\providecommand{\doi}[1]{https://doi.org/#1}

\bibitem{australian2001}
{Australian Screening Advisory Committee}: {National Accreditation Standards BreastScreen Australia Quality Improvement Program (Revised)} (2001)

\bibitem{brahim2022automated}
Brahim, M., Westerkamp, K., Hempel, L., Lehmann, R., Hempel, D., Philipp, P.: Automated assessment of breast positioning quality in screening mammography. Cancers  \textbf{14}(19), ~4704 (2022)

\bibitem{globalcancer}
{Cancer (IARC), T. I. A. for R. on Global Cancer Observatory}: {Global Cancer Observatory}. \url{https://gco.iarc.fr/}, accessed on 14 May 2024

\bibitem{duffy2002impact}
Duffy, S.W., Tab{\'a}r, L., Chen, H.H., Holmqvist, M., Yen, M.F., Abdsalah, S., Epstein, B., Frodis, E., Ljungberg, E., Hedborg-Melander, C., et~al.: The impact of organized mammography service screening on breast carcinoma mortality in seven swedish counties: a collaborative evaluation. Cancer: Interdisciplinary International Journal of the American Cancer Society  \textbf{95}(3),  458--469 (2002)

\bibitem{feig2002}
Feig, S.A.: {Image quality of screening mammography: effect on clinical outcome}. AJR Am J Roentgenol  \textbf{178},  805--807 (2002)

\bibitem{feng2018wing}
Feng, Z.H., Kittler, J., Awais, M., Huber, P., Wu, X.J.: Wing loss for robust facial landmark localisation with convolutional neural networks. In: Proceedings of the IEEE conference on computer vision and pattern recognition. pp. 2235--2245 (2018)

\bibitem{geras2019artificial}
Geras, K.J., Mann, R.M., Moy, L.: Artificial intelligence for mammography and digital breast tomosynthesis: current concepts and future perspectives. Radiology  \textbf{293}(2),  246--259 (2019)

\bibitem{gupta2020deep}
Gupta, V., Taylor, C., Bonnet, S., Prevedello, L.M., Hawley, J., White, R.D., Flores, M.G., Erdal, B.S.: Deep learning-based automatic detection of poorly positioned mammograms to minimize patient return visits for repeat imaging: A real-world application. arXiv preprint arXiv:2009.13580  (2020)

\bibitem{gurdemir2012}
Gürdemir, B., Arıbal, E.: {Assessment of mammography quality in Istanbul}. Diagn Interv Radiol  \textbf{18},  468--472 (2012)

\bibitem{hendrick1999}
Hendrick, R.E., Bassett, L., Botsco, M.A., et~al.: {Mammography quality control manual}. Royal American College of Radiologists (1999)

\bibitem{mackenzie2016}
Mackenzie, A., Warren, L.M., Wallis, M.G., Given-Wilson, R.M., Cooke, J., Dance, D.R., Chakraborty, D.P., Halling-Brown, M.D., Looney, P.T., Young, K.C.: The relationship between cancer detection in mammography and image quality measurements. Physica Medica  \textbf{32}(4),  568--574 (2016)

\bibitem{magnus2011effectiveness}
Magnus, M.C., Ping, M., Shen, M.M., Bourgeois, J., Magnus, J.H.: Effectiveness of mammography screening in reducing breast cancer mortality in women aged 39--49 years: a meta-analysis. Journal of women's health  \textbf{20}(6),  845--852 (2011)

\bibitem{nguyen2023vindr}
Nguyen, H.T., Nguyen, H.Q., Pham, H.H., Lam, K., Le, L.T., Dao, M., Vu, V.: Vindr-mammo: A large-scale benchmark dataset for computer-aided diagnosis in full-field digital mammography. Scientific Data  \textbf{10}(1), ~277 (2023)

\bibitem{rodriguez2019stand}
Rodriguez-Ruiz, A., L{\aa}ng, K., Gubern-Merida, A., Broeders, M., Gennaro, G., Clauser, P., Helbich, T.H., Chevalier, M., Tan, T., Mertelmeier, T., et~al.: Stand-alone artificial intelligence for breast cancer detection in mammography: comparison with 101 radiologists. JNCI: Journal of the National Cancer Institute  \textbf{111}(9),  916--922 (2019)

\bibitem{rancr2002}
{Royal Australian and New Zealand College of Radiologists}: {Mammography quality assurance program} (2002)

\bibitem{spuur2011mammography}
Spuur, K., Hung, W.T., Poulos, A., Rickard, M.: Mammography image quality: model for predicting compliance with posterior nipple line criterion. European journal of radiology  \textbf{80}(3),  713--718 (2011)

\bibitem{fda2016}
{U.S. Food and Drug Administration}: {Positioning Responsible For Most Clinical Image Deficiencies, Failures}. \url{https://www.fda.gov/Radiation-EmittingProducts/MammographyQualityStandardsActandProgram/FacilityScorecard/ucm495378.html} (2016), accessed on 14 May 2024

\bibitem{watanabe2023quality}
Watanabe, H., Hayashi, S., Kondo, Y., Matsuyama, E., Hayashi, N., Ogura, T., Shimosegawa, M.: Quality control system for mammographic breast positioning using deep learning. Scientific Reports  \textbf{13}(1), ~7066 (2023)

\bibitem{wilson2011}
Wilson, R., Liston, J.: {Quality assurance guidelines for radiographers}. NHSBSP Publication, 2nd edn. (2011)

\end{thebibliography}

\end{document}


\title{Appendix}


%
%


\author{\textit{Anonymized}}




%
\authorrunning{Tanyel et al.}
%
%
\maketitle              

\begin{figure}[htb!]
\begin{center}
\resizebox{\textwidth}{!}{
\includegraphics[]{figures/miccai-flow.drawio.png}}
\caption{A specialized browser-based platform was used by an expert radiologist to annotate the pectoral line and nipple. The posterior nipple line is automatically drawn using the perpendicular rule. After preprocessing the images, the proposed deep learning methodology was applied, with models trained on an 80\%/10\%/10\% split for training, validation, and testing. The models' performance was evaluated on a separate test set, using the posterior nipple line criterion to distinguish between good and poor positioning in mediolateral oblique views by predicting anatomical landmarks.} \label{Fig1}
\end{center}
\end{figure}

\section{Detailed Methodology}
\subsection{Study Sample}
For this study, we utilized the publicly available VinDr Mammography dataset \cite{nguyen2023vindr}, which is an extensive open-access collection comprising 5000 mammography exams. These exams were gathered from opportunistic screening settings at two hospitals in Vietnam between 2018 and 2020.

To make the most of our available resources, we selected 1000 exams from the 5000 available. Each exam consists of two MLO view mammograms from both breasts, resulting in a total of 2000 mammography images in our dataset. These 2000 images were then divided into a development set (comprising training and validation subsets) and a separate testing set. We used an 80\%/10\%/10\% split for this division, which was randomized to ensure a balanced representation of various clinical outcomes within the study.

\subsection{Image Positioning Quality Criterion}
Several international systems have been developed to evaluate the quality of mammography images, each proposing different criteria for assessing the positioning in MLO views. These criteria often include aspects such as the angle, width, and length of the pectoral muscle, the angulation of its anterior and posterior borders, the distance between the inferior aspect of the pectoral muscle and the nipple level, and the PNL extending from the nipple to the pectoralis muscle.

Despite these varied criteria, the central objective is to ensure maximum breast tissue coverage in the MLO view. Some criteria, like the distance from the inferior aspect of the pectoral muscle to the nipple level, are highly subjective and frequently impractical to achieve \cite{spuur2011mammography}. Others, such as the angle and dimensions of the pectoral muscle, do not have universally accepted standards for optimal positioning.

A universally applicable and consistent criterion is that the PNL, when drawn from the nipple to the pectoralis muscle at a right angle, should intersect the pectoralis muscle rather than just reaching the edge of the film. This method is endorsed by both the American College of Radiology and the Royal Australian and New Zealand College of Radiologists \cite{hendrick1999, australian2001, rancr2002, wilson2011, spuur2011mammography}. Consequently, we have adopted this criterion as the reference standard for our current study.

\subsection{Preprocessing Steps}
The preprocessing steps for the mammography images are designed to prepare the data for subsequent analysis while preserving important anatomical features and spatial relationships. The following steps outline the preprocessing methodology:

\begin{enumerate}
    \item \textbf{Landmark Extraction:} The midpoint of the nipple bounding boxes and the endpoints of the pectoralis muscle are extracted from the images. This process yields three critical landmarks for each image, which are crucial for orientation and scale adjustments in further processing steps.

    \item \textbf{Standardization of Pectoralis Muscle Landmarks:} The endpoints of the pectoralis muscle are standardized to minimize the variability introduced by the arbitrary termination of the muscle lines by radiologists. Given the curvature of the muscle and the radiologists' focus on covering the muscle adequately, the endpoint of each line is extended to the image boundary with a consistent margin of 10 pixels. This adjustment ensures that edge-related errors do not impact subsequent analyses.

    \item \textbf{Cropping:} Significant regions of the breast are isolated by removing extensive black pixel areas typically present around the periphery of the images and below the breast area. This step focuses the analysis on relevant breast tissue and reduces computational load. The cropping process involves calculating the threshold value as the mean of the image, converting the image to a binary image based on this threshold, applying morphological opening with a disk-shaped structuring element to clean the binary image, labeling connected regions in the cleaned binary image, identifying and selecting the largest region based on area, and finally extracting the bounding box of this largest region and cropping the image accordingly.

    \item \textbf{Padding:} To prevent distortion during resizing and to maintain the high quality of the mammograms, zero-padding is applied to make the images square. This step is crucial for maintaining the aspect ratio and ensuring uniformity across the dataset.

    \item \textbf{Resizing:} All images are resized to a standard dimension of $512 \times 512$ pixels. This resizing facilitates the computational efficiency of subsequent processing steps and model training without sacrificing detail necessary for accurate analysis.

    \item \textbf{Alignment and Scaling of Landmarks:} Throughout the preprocessing steps, careful attention is given to aligning the landmarks' coordinates with the modifications made to the images. Additionally, pixel spacing information is meticulously preserved, enabling the recovery of real-world measurements (in millimeters) and ensuring that spatial relationships within the images remain accurate.
\end{enumerate}

\subsection{Evaluation}

The evaluation focuses on assessing the accuracy of our deep learning model in predicting three key landmarks: the nipple and the endpoints of the pectoral muscle line. We aim to verify whether the perpendicular intersection of the line drawn from the nipple to the pectoral muscle is correctly predicted and falls within the image boundaries, thereby indicating good or poor image quality.

In addition to this, we calculate the angular error between the original and predicted pectoral muscle lines to gauge the model's ability to predict the line's orientation accurately. Furthermore, we evaluate model performance by computing accuracy, sensitivity, and specificity for both the automatic decision-making system based on the perpendicular point's location and second-opinion labels provided by another radiologist. This comparison provides a more intuitive and clinical assessment of the model's reliability in practical scenarios.

\subsubsection{Perpendicular Point Calculation}
The function $\mathcal{P}(x_1, y_1, x_2, y_2, x_n, y_n)$ calculates the intersection point of a perpendicular line drawn from the nipple coordinates \((x_n, y_n)\) to the pectoral muscle line, which is defined by the endpoints \((x_1, y_1)\) and \((x_2, y_2)\). The function $\mathcal{P}$ efficiently processes lines with arbitrary slopes by calculating the slope and intercept of the perpendicular line and analytically determining the coordinates of the intersection.

\subsubsection{Image Boundary Check}
To assess the predicted landmarks' quality, we ensure that the perpendicular intersection falls within the image boundaries using the function $\mathcal{I}(\mathcal{S}, p)$, where $\mathcal{S}$ represents the shape of the image, and \(p\) denotes the intersection point's coordinates. If the point is within bounds, the image is considered of good quality; otherwise, it is classified as poor quality.

\subsubsection{Millimeter Distance Calculation}
After validating the image quality based on the perpendicular intersection point, we compute the Euclidean distance between the predicted and original landmarks in millimeters using the function $\mathcal{D}(p_a, p_b, d_f)$. Here, $p_a$ represents the predicted landmark position (e.g., the predicted nipple position or one endpoint of the predicted pectoral muscle line), while $p_b$ represents the corresponding original landmark position from the ground truth data. The parameter $d_f$ represents a row from metadata associated with the image or the landmarks; specifically, it holds the pixel spacing information adjusted during preprocessing to convert back to the real-world millimeter distance.

\subsubsection{Angle Error Calculation}
We evaluate the angle error between the original and predicted pectoral muscle lines using the function $\mathcal{A}(x_1, y_1, x_2, y_2, \text{orientation})$. This function calculates the angle of a line with respect to the vertical, considering the side-specific orientation of the breast (left or right). The calculated angle represents how much the predicted line deviates from being parallel to the vertical side of the breast. This angle is normalized to a range of 0 to 180 degrees to quantify the error in the orientation of the predicted pectoral muscle line with respect to the original line, providing a geometrically intuitive measure of model accuracy.

\subsubsection{Model Performance Metrics}
We further assess the performance of our model using standard statistical metrics: accuracy, sensitivity, and specificity. These metrics are derived based on the 'good' or 'bad' quality classification determined by the position of the perpendicular intersection point relative to the image boundaries. Accuracy measures the proportion of total predictions (both good and bad) that were correctly identified by the model. Sensitivity measures the proportion of actual bad quality images that were correctly identified as bad by the model. Specificity measures the proportion of actual good quality images that were correctly identified as good by the model.

\subsection{Model Architectures and Techniques for Landmark Detection}

\begin{figure}[htb!]
\begin{center}
\resizebox{\textwidth}{!}{
\includegraphics[]{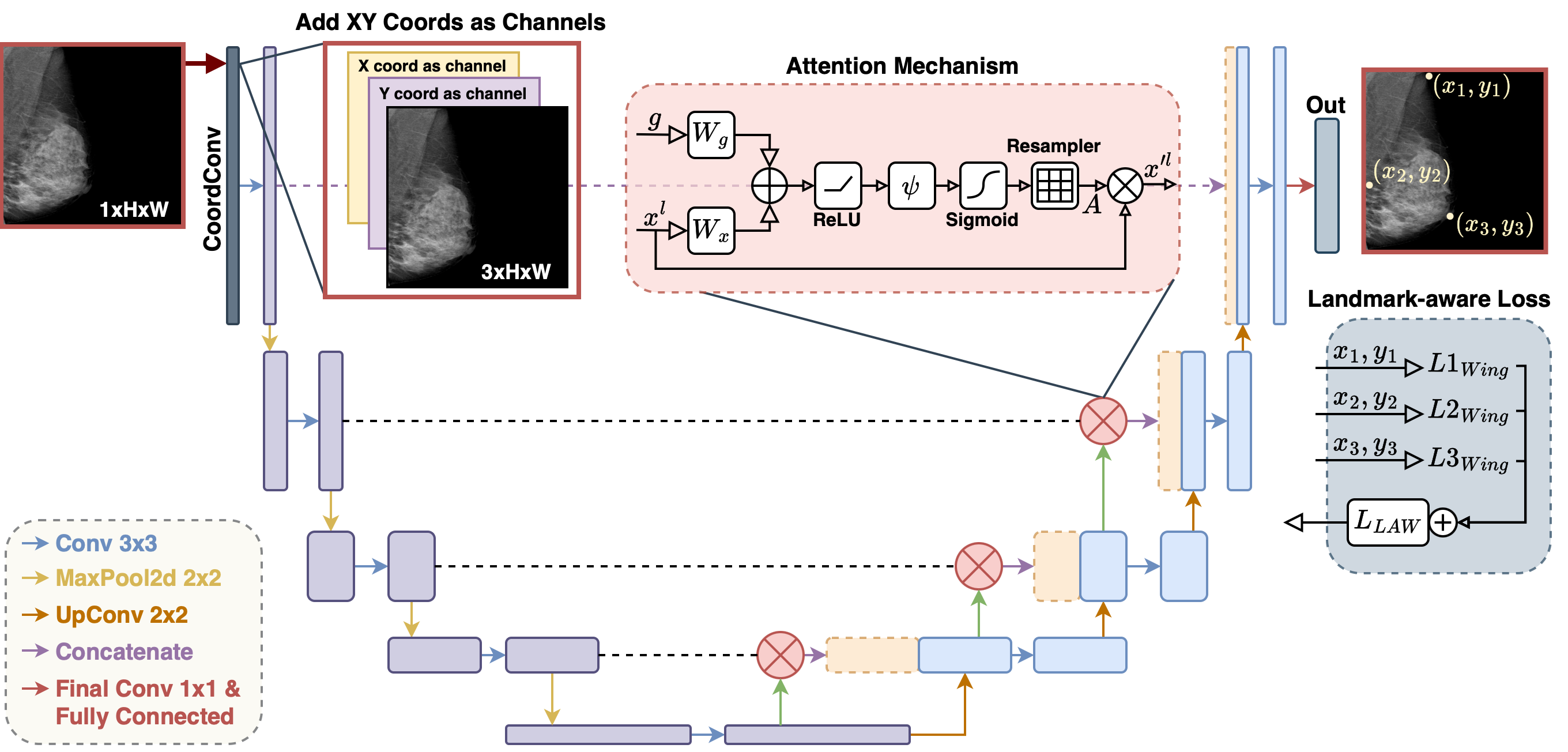}}
\caption{Illustration of concepts utilized in this study as part of an ablation study. Our method integrates CoordConv layer, an attention mechanism, and landmark-aware loss. At the input layer, a single-channel (1xHxW) grayscale mammogram is augmented to a three-channel (3xHxW) image by introducing two additional channels that encode the X and Y spatial coordinates of each pixel. The attention mechanism refines features, and skip connections preserve spatial information. The final layer outputs landmark coordinates, optimized using landmark-aware loss.} \label{Fig2}
\end{center}
\end{figure}

The proposed methodology leverages U-Net as the backbone, incorporating coordinate convolutions (CoordConv) and attention mechanisms as enhancement methods for landmark detection (Fig. \ref{Fig2}). By integrating CoordConv and attention mechanisms, the feature maps are refined throughout the network. Additionally, we utilized ResNeXt50 as a backbone to design a landmark regressor using the same final layer, comparing the raw classification model of ResNeXt50 to image-level classification and regression results. This section details the foundations of these components and the landmark regression process, which are crucial for precise landmark detection.

\subsubsection{CoordConv Integration}

To improve spatial awareness, the initial convolutional layer in the encoder, denoted as $\mathcal{C}_1$, is replaced by a CoordConv layer in CoordAtt UNet. This approach integrates spatial coordinates directly into the convolutional operation, enabling the network to learn spatial hierarchies more effectively. The CoordConv layer, $\mathcal{CC}$, augments the input feature map by appending normalized spatial coordinates across the height ($H$) and width ($W$) dimensions of the input tensor $I$. The transformation is defined as:

\[
\mathcal{CC}(I) = \mathcal{R} \left( \mathcal{B} \left( \mathcal{C} \left( \mathcal{A}(I, \{x, y\}) \right) \right) \right),
\]

where:
\begin{itemize}
    \item $\mathcal{A}(I, \{x, y\})$ represents the addition of coordinate channels to the input tensor $I$. The x-coordinates $\{x\}$ and y-coordinates $\{y\}$ are normalized in the range [0, 1], calculated as:
    \[
    x_j = \frac{j}{W-1}, \quad y_i = \frac{i}{H-1},
    \]
    for each pixel $(i, j)$ in the feature map. These coordinates are repeated across the batch size and stacked to form two new channels appended to $I$.
    
    \item $\mathcal{C}$ denotes a convolution operation performed using the expanded input channels (original input channels plus two coordinate channels). This operation enhances the model's ability to leverage positional information.
    
    \item $\mathcal{B}$ and $\mathcal{R}$ represent batch normalization and a ReLU activation function, respectively, which normalize the outputs and introduce non-linearity.
\end{itemize}

\subsubsection{Attention Mechanism}

Attention mechanisms in the U-Net model selectively focus on important features. The attention block integrates features from both encoder and decoder paths by computing attention coefficients (\(\psi\)) using the following formulations: \(g = W_{\text{gate}} \ast G\) and \(x = W_{x} \ast X\). Here, \(W_{\text{gate}}\) and \(W_{x}\) are convolutional filters, with \(G\) representing the gating signal from the decoder and \(X\) the feature map from the encoder. The attention coefficients \(\psi\) are computed as \(\psi = \sigma(\text{ReLU}(g + x))\), where \(\sigma\) denotes the sigmoid activation. The attended output \(A\) is then calculated as \(A = X \cdot \psi\). This mechanism ensures that only the most relevant features are propagated through the network to improve the precision of the output.

\subsubsection{Landmark Regression}

For landmark regression, the high-level feature maps are processed through a 1x1 convolution layer to reduce the channel dimension to the desired number of output features. This is followed by adaptive average pooling to a fixed size, after which the pooled features are flattened. Finally, a fully connected layer produces the landmark coordinates. This compact process ensures efficient and accurate prediction of landmark positions, leveraging high-level feature representations refined by attention mechanisms.

\subsubsection{ResNeXt50}

We also utilized ResNeXt50 as part of our study. The ResNeXt50 architecture features a 50-layer deep residual network with a cardinality of 32, allowing for enhanced feature learning through multiple pathways. The model utilizes grouped convolutions, effectively expanding the network's capacity to learn diverse feature representations.

For our landmark regression task, we modified the ResNeXt50 model by adjusting the first convolution layer to accept single-channel input and the final layer to output the required number of landmark coordinates. Specifically, the number of output features was set to match the number of landmarks (e.g., six for three landmarks with x and y coordinates).

The raw classification model of ResNeXt50 was also utilized to compare image-level classification to regression results. This comparison allowed us to evaluate the differences in accuracy and robustness between the two approaches, demonstrating the versatility and effectiveness of the ResNeXt50 architecture in handling various types of tasks within the same framework.

\bibliographystyle{splncs04}
\bibliography{mybib}